\documentclass[10pt,conference]{IEEEtran}
\IEEEoverridecommandlockouts
\usepackage{cite}
\usepackage{amsmath,amssymb,amsfonts}
\usepackage{algorithmic}
\usepackage{graphicx}
\usepackage{textcomp}
\usepackage{xcolor}

\def\BibTeX{{\rm B\kern-.05em{\sc i\kern-.025em b}\kern-.08em
    T\kern-.1667em\lower.7ex\hbox{E}\kern-.125emX}}
\begin{document}

\title{Zero-Shot Knowledge Base Resizing for Rate-Adaptive Digital Semantic Communication
}

\author{\IEEEauthorblockN{Shumin Yao$^{1}$, Hui Du$^{1}$, Lifeng Xie$^{1,*}$, Yaping Sun$^{1}$, Hao Chen$^{1}$, Nan Ma$^{2, 1}$, ~\IEEEmembership{Member, IEEE}, \\ and Xiaodong Xu$^{2, 1}$,~\IEEEmembership{Senior Member, IEEE}}
\IEEEauthorblockA{$^{1}$Department of Broadband Communication, Pengcheng Laboratory, Shenzhen 518066, China \\
$^{2}$State Key Laboratory of Networking
and Switching Technology,  Beijing University of Posts and \\ Telecommunications, Beijing 100876, China\\
\{yaoshm, duh, xielf,  sunyp, chenh03\}@pcl.ac.cn, \{manan, xuxiaodong\}@bupt.edu.cn}}

\maketitle

\begin{abstract}
Digital semantic communication systems, which often leverage the Vector Quantized Variational Autoencoder (VQ-VAE) framework, are pivotal for future wireless networks. In a VQ-VAE-based semantic communication system, the transmission rate is directly governed by the size of a discrete codebook known as knowledge base (KB). However, the KB size is a fixed hyperparameter, meaning that adapting the rate requires training and storing a separate model for each desired size—a practice that is too computationally and storage-prohibitive to achieve truly granular rate control. To address this, we introduce a principled, zero-shot KB resizing method that enables on-the-fly rate adaptation without any retraining. Our approach establishes a global importance ranking for all vectors within a single, large parent KB by uncovering its inherent semantic hierarchy. This is achieved via a three-step framework: 1) embedding KB vectors into hyperbolic space to reveal their hierarchical relationships; 2) constructing a master semantic tree using a minimum spanning tree algorithm; 3) enabling instant resizing by iteratively pruning the least important leaf nodes. Extensive simulations demonstrate that our method achieves reconstruction quality nearly identical to that of dedicated KBs trained from scratch, while demanding only a fraction of the computational budget. Moreover, our approach exhibits superior robustness at very low rates, where conventional KBs suffer from catastrophic failure. Our work resolves a fundamental limitation of VQ-VAE-based semantic communication systems, offering a practical and efficient path toward flexible and rate-adaptive semantic communication.
\end{abstract}


\section{Introduction}

Semantic communication is emerging as a new paradigm poised to move beyond traditional bit-level fidelity by optimizing the transmission of meaning and relevance, a crucial step for addressing spectrum scarcity in future wireless networks \cite{ren2024knowledge}. To ensure compatibility with existing digital infrastructure, recent research has focused on digital semantic communication systems, typically built upon the vector quantized variational autoencoder (VQ-VAE) framework \cite{vqvae}. This framework establishes a shared discrete semantic knowledge base (KB), or codebook, containing a finite set of representative vectors. After an encoder extracts a tensor of continuous semantic features from the source message, a process known as vector quantization is performed: each feature vector in the tensor is individually mapped to a discrete index corresponding to a specific vector in the KB that serves as its closest match. A decoder at the receiver then uses these transmitted indices to retrieve the corresponding vectors and reconstruct the original message \cite{RobustVQSemCom}.

This VQ-based approach, which relies on transmitting indices, enforces a critical trade-off directly governed by the knowledge base size, defined as the total number of vectors, $K$. On one hand, a large KB with a high $K$ value enables a finer-grained semantic representation, which minimizes quantization error and significantly improves reconstruction quality. On the other hand, this fidelity directly increases the transmission rate, as representing each index requires $\lceil \log_2 K \rceil$ bits. A larger $K$ therefore increases the total data payload, demanding greater channel bandwidth. This inherent conflict means the optimal KB size is not static; it should be dynamically adapted to balance reconstruction quality with available channel resources. However, such adaptability is fundamentally incompatible with the VQ-VAE framework's design. The KB size, $K$, is a fixed hyperparameter determined prior to training, and this architectural rigidity is an inherent limitation passed down to the entire class of digital semantic communication systems built upon VQ-VAE (e.g., \cite{RobustVQSemCom, VQSemCom}). Consequently, adapting the transmission rate by changing $K$ would require training and storing a separate, complete model for each potential size—a practice that is computationally and storage-prohibitive. 

To address this limitation, some works have proposed to approximate the flexible rate adaptation by selecting from a discrete set of pre-defined KB configurations. For instance, some methods achieve different rates by varying the number of active sub-KBs within a single large KB \cite{VPQSemCom} or by using a variable number of its hierarchical layers \cite{MOC-RVQ}. Another approach involves training multiple distinct KBs, each tailored for a different point on the rate-distortion curve, and then switching between them during inference \cite{ESC-MVQ}. However, this reliance on a discrete set of pre-defined KB configurations, meaning the system is fundamentally constrained to a small, fixed number of rate levels. While a recent retraining-free resizing technique has been proposed \cite{HyperHill}, its core mechanism is a computational shortcut that fails to reduce the actual transmission rate. Thus, the ability to restructure a KB to an arbitrary size for truly granular, on-the-fly adaptation remains a significant open problem.

In this paper, we propose a novel, zero-shot KB resizing method that enables truly rate-adaptive semantic communication. Our core idea lies within establishing a global importance ranking for all vectors in a large KB, allowing us to form a new, smaller KB by simply retaining a desired number of the most important vectors. Here, the global importance ranking is established by exploiting the principle that the semantic relationships within a well-trained KB are inherently hierarchical. Our contributions are summarized as follows.
\begin{enumerate}
    \item We propose a complete, three-step framework for zero-shot KB resizing. The method first leverages hyperbolic embeddings to reveal the inherent hierarchy of KB vectors, then constructs a master semantic tree via a minimum spanning tree (MST) to make this hierarchy explicit, and finally enables instant resizing to any target size by iteratively pruning the tree's leaf nodes.
    \item We perform extensive simulations to demonstrate the effectiveness of our approach. The results show that our zero-shot resizing method achieves reconstruction quality nearly identical to that of dedicated models trained from scratch across a wide range of KB sizes, while demanding only a fraction of the required training budget. Moreover, we reveal that our method provides superior robustness and graceful quality degradation at very low rates, where the baseline approach suffers from catastrophic failure.
\end{enumerate}

The rest of the paper is organized as follows. Section~\ref{sec:system_model} details the VQ-VAE-based system model for digital semantic communication. Section~\ref{sec:resizing_method} presents our proposed zero-shot KB resizing method, including its three-step framework. In Section~\ref{sec:Exp}, we evaluate the performance of our approach through extensive simulations. Finally, Section~\ref{sec:Conclusion} concludes the paper.

\section{System Model}
\label{sec:system_model}

\begin{figure}
    \centering
    \includegraphics[width=1\linewidth]{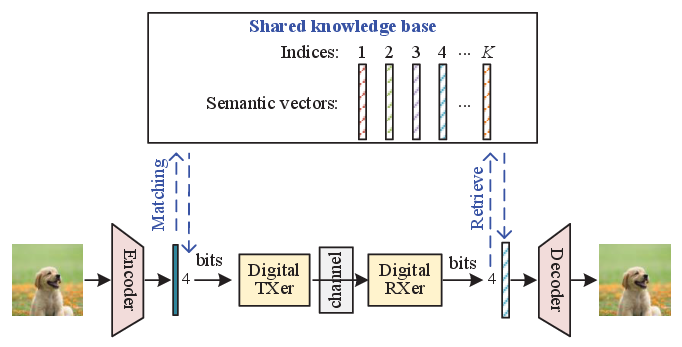}
    \caption{Considered VQ-VAE-based digital semantic communication framework.}
    \label{fig:VQVAEFramework}
\end{figure}

This work considers a semantic communication system deployed on digital hardware, where information is ultimately processed and transmitted as binary bits. The system architecture is built upon the VQ-VAE framework \cite{vqvae}, a representative model for end-to-end digital semantic information transmission. As illustrated in Fig.~\ref{fig:VQVAEFramework}, the system comprises a sender and a receiver that share a common, pre-trained semantic knowledge base (SKB).

Specifically, the sender employs a deep neural network (DNN) based semantic encoder, $\mathcal{E}_{\mathbf{\theta}_{\text{enc}}}(\cdot)$, with trainable parameters $\mathbf{\theta}_{\text{enc}}$. The receiver is equipped with a corresponding DNN-based semantic decoder, $\mathcal{D}_{\mathbf{\theta}_{\text{dec}}}(\cdot)$, with parameters $\mathbf{\theta}_{\text{dec}}$. A shared SKB, denoted as a set of $K$ embedding vectors $\mathcal{S} = \{\mathbf{s}_k\}_{k=1}^{K}$, where each $\mathbf{s}_k \in \mathbb{R}^{C'}$, is stored at both ends.

For clarity, we focus on an image transmission scenario. Let $\mathbf{I} \in \mathbb{R}^{H \times W \times C}$ be a source image. The sender first uses the semantic encoder to extract a tensor $\mathbf{X} \in \mathbb{R}^{H' \times W' \times C'}$:
\begin{equation}
    \label{eq:sem_enc}
    \mathbf{X} = \mathcal{E}_{\mathbf{\theta}_{\text{enc}}}(\mathbf{I}).
\end{equation}
Here, $H' < H$ and $W' < W$, signifying semantic compression of the spatial dimensions.

Next, for each vector $\mathbf{x}_{h',w'} \in \mathbb{R}^{C'}$ in the tensor $\mathbf{X}$, the sender performs vector quantization. It finds the nearest vector in $\mathcal{S}$ using the $\ell_2$-norm distance and records its index:
\begin{equation}
    \label{eq:quantize}
    z_{h',w'} = \underset{k \in \{1, \dots, K\}}{\operatorname{argmin}} ||\mathbf{x}_{h',w'} - \mathbf{s}_{k}||_{2}.
\end{equation}
This process maps the continuous feature tensor $\mathbf{X}$ to a discrete integer tensor of indices $\mathbf{Z}_{\text{idx}} \in \mathbb{Z}^{H' \times W'}$, where each element is an integer from $1$ to $K$. This quantization process is what makes the system inherently digital.

The tensor of indices, $\mathbf{Z}_{\text{idx}}$, is then transmitted to the receiver. For transmission, each index $z_{h',w'}$ is represented by a binary codeword of length $B$ bits, where
\begin{equation}
    \label{eq:bit_num}
    B = \lceil \log_{2}{K} \rceil.
\end{equation}
The total transmission overhead for the image is thus $H' \times W' \times B$ bits, neglecting any further channel coding.

Upon receiving $\mathbf{Z}_{\text{idx}}$ (which we assume to be error-free for the purpose of defining the model), the receiver performs a lookup operation. It reconstructs the quantized tensor, $\mathbf{X}_{q} \in \mathbb{R}^{H' \times W' \times C'}$, by replacing each index $z_{h',w'}$ with its corresponding vector from the shared SKB: $(\mathbf{X}_{q})_{h',w'} = \mathbf{s}_{z_{h',w'}}$. Finally, the semantic decoder uses this quantized tensor to reconstruct the source image:
\begin{equation}
    \label{eq:sec_dec}
    \mathbf{\hat{I}} = \mathcal{D}_{\mathbf{\theta}_{\text{dec}}}(\mathbf{X}_{q}).
\end{equation}

The size of the KB, $K$, governs a fundamental trade-off in system performance. A larger $K$ enables finer-grained semantic representation, reducing quantization error and improving reconstruction quality of $\mathbf{\hat{I}}$. However, this enhanced fidelity incurs substantial costs. First, the transmission payload increases with $B = \lceil \log_{2}{K} \rceil$ bits per index, demanding greater channel resources in terms of bandwidth or transmission time. Second, the computational burden at the sender escalates, as the nearest-neighbor search in \eqref{eq:quantize} requires $O(K)$ distance calculations per feature vector, resulting in higher processing latency and energy consumption. Conversely, a smaller $K$ reduces both transmission and computational overhead but at the expense of coarser semantic quantization and degraded reconstruction quality. The ideal value of $K$ is therefore not fixed, but depends on dynamic factors like channel conditions and hardware constraints. Unfortunately, existing VQ-VAE frameworks lack this capability: conventional approaches require training and storing separate models for each possible KB size, rendering them impractical due to prohibitive storage and computational costs. Addressing this fundamental limitation motivates the zero-shot resizing method presented in this work.

\begin{figure*}
    \centering
    \includegraphics[width=0.85\linewidth]{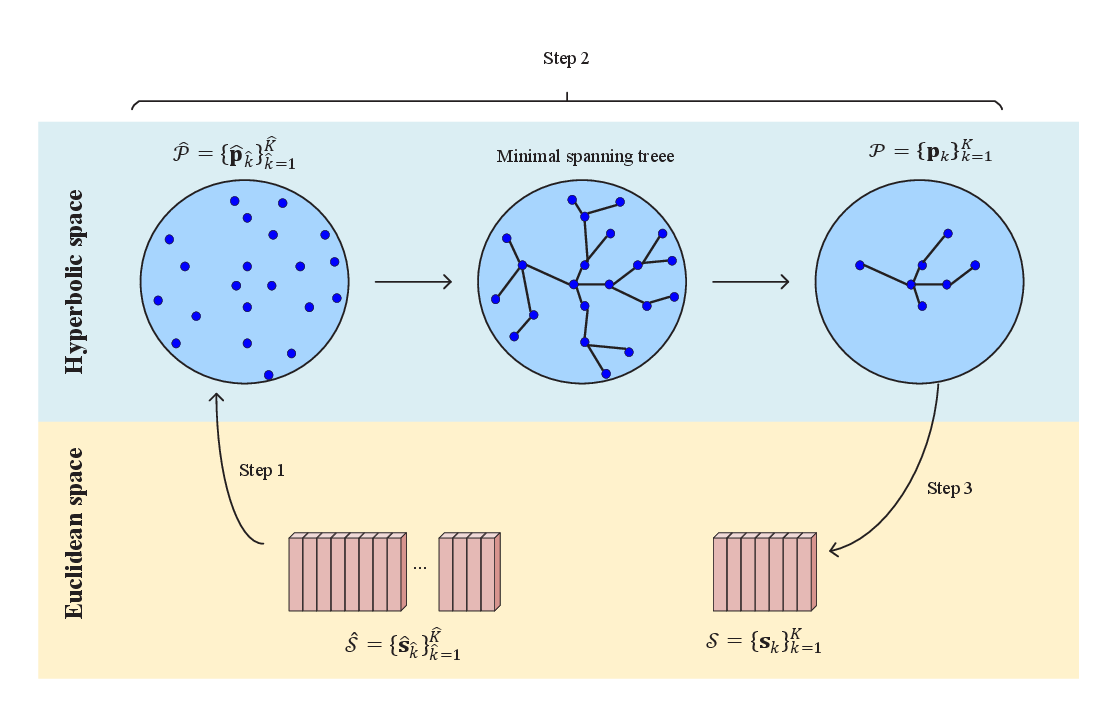}
    \caption{Procedures of the proposed zero-shot knowledge base resizing}
    \label{fig:steps}
\end{figure*}

\section{Zero-Shot Knowledge Base Resizing}
\label{sec:resizing_method}

This section details our proposed zero-shot KB resizing method. Our core idea is to select a principled subset of vectors from a single, large, and well-trained KB, allowing it to be dynamically adapted to any smaller size as the communication system demands—all without requiring any costly retraining or re-optimization. The central challenge, therefore, lies in making this selection in a way that ensures the resulting smaller KB covers knowledge as comprehensively as possible, so that it can support semantic communication as effectively as possible.

To address this challenge, we exploit the principle that the semantic relationships within a well-trained KB are inherently hierarchical. This principle reflects a fundamental property of semantic information, where broad categories naturally contain more specific concepts, a structure evident in both language and vision \cite{khrulkov2020hyperbolic}. This hierarchy allows us to perform a principled resizing: we prioritize preserving the coarse-grained, foundational concepts that define the overall knowledge structure, and then include as many fine-grained, specific concepts as the target size permits. To achieve this, as shown in Fig. \ref{fig:steps}, our method performs three key steps. First, we embed the vectors of a large, pre-trained KB (of size $\hat{K}$) from their native Euclidean space into a hyperbolic space, a space whose geometry is uniquely suited for representing hierarchical structures with minimal distortion. Second, we construct a master semantic tree by building a MST based on hyperbolic distances, which enables zero-shot resizing to any target size $K < \hat{K}$ through the deterministic pruning of this tree. Finally, the vectors of the pruned tree are mapped back from hyperbolic space to the original Euclidean space to produce the final, ready-to-use KB. The remainder of this section details these three steps.

\subsection{Hyperbolic embedding for revealing semantic granularity}

Let $\hat{\mathcal{S}}=\left\{\hat{\mathbf{s}}_{\hat{k}}\right\}_{\hat{k}=1}^{\hat{K}}$ denote the set of vectors in the large, pre-trained KB. Our first step is to reveal the semantic hierarchy within $\hat{\mathcal{S}}$ by embedding its vectors into a space that naturally represents such structures. We posit that the relationships between semantic concepts are inherently hierarchical, a property for which hyperbolic geometry provides a more faithful representation with lower distortion than Euclidean space \cite{khrulkov2020hyperbolic, nickel2017poincare}. By mapping the KB into the Poincaré ball model, we create a geometric arrangement where each vector's position reflects its corresponding semantic granularity: vectors with coarse-grained, foundational semantics are placed near the center of the ball, while vectors with fine-grained, specific semantics are pushed towards its boundary.

The embedding is achieved via the exponential map, which projects each Euclidean vector $\hat{\mathbf{s}}_{\hat{k}}$ from the origin $\mathbf{0}$ onto the Poincaré ball. For a ball with curvature $c=1$, the mapping is defined as:
\begin{equation}
    \label{eq:exp_map}
    \hat{\mathbf{p}}_{\hat{k}} = \tanh\left(\left\|\hat{\mathbf{s}}_{\hat{k}}\right\|\right) \frac{\hat{\mathbf{s}}_{\hat{k}}}{\left\|\hat{\mathbf{s}}_{\hat{k}}\right\|}, \quad \text{for } \hat{k} = 1, \dots, \hat{K}.
\end{equation}
This transformation produces the set of hyperbolic vectors $\hat{\mathcal{P}}=\left\{\hat{\mathbf{p}}_{\hat{k}}\right\}_{\hat{k}=1}^{\hat{K}}$, where each vector's distance from the origin now serves as a proxy for its semantic importance. This provides the geometric foundation required for constructing an explicit semantic tree in the next step.

\subsection{Hierarchical pruning for zero-shot resizing}

The set of hyperbolic vectors $\hat{\mathcal{P}}$ produced in the previous step is a point cloud geometrically organized by semantic granularity. The goal of this step is to transform this point cloud into an explicit hierarchical structure—a master semantic tree—which can then be deterministically pruned to any desired size $K$.

Our resizing strategy is based on a core principle: we must preserve the foundational, coarse-grained vectors that form the skeleton of the KB, while controllably pruning the more specific, fine-grained vectors. This requires a structure that explicitly defines the parent-child relationships between vectors. An MST is the ideal mathematical tool for this task; it is a principled, parameter-free method for discovering the most likely underlying structure in a set of points. By connecting all vectors using the shortest possible total path length, the MST reveals the most likely semantic hierarchy latent within the geometry of the point cloud.

The MST is constructed using the hyperbolic distance, $d_{\mathcal{H}}$, as the edge weight connecting any two vectors $\mathbf{p}_i$ and $\mathbf{p}_j$:
\begin{equation}
    \label{eq:hyperbolic_dist}
    d_{\mathcal{H}}(\mathbf{p}_i, \mathbf{p}_j) = \operatorname{arccosh}\left(1 + 2 \frac{\|\mathbf{p}_i - \mathbf{p}_j\|^2}{(1 - \|\mathbf{p}_i\|^2)(1 - \|\mathbf{p}_j\|^2)}\right).
\end{equation} 

We construct the master semantic tree using Prim's algorithm \cite{prim_alg}. We designate the vector with the smallest hyperbolic distance to the origin of the Poincaré ball as the root of the tree, as it represents the most general concept in the entire KB. The algorithm then iteratively adds the nearest unconnected vector to the growing tree until all $\hat{K}$ vectors are connected.

This master tree, containing all $\hat{K}$ nodes, serves as a blueprint for resizing. To obtain a KB of a specific size $K < \hat{K}$, we perform a deterministic pruning process. We iteratively remove the leaf nodes from the current tree until exactly $K$ nodes remain. We denote this final set of $K$ vectors as $\mathcal{P}_K$. Because leaf nodes in the semantic tree represent the most fine-grained concepts, this iterative process directly realizes our core principle of preserving the foundational, coarse-grained skeleton of the KB while systematically removing the most specific vectors. While more complex pruning strategies based on centrality or other metrics could be devised, this leaf-based approach is selected for its computational simplicity and deterministic nature. Moreover, this process can be performed instantly for any target size $K$, fulfilling the promise of zero-shot resizing.

\subsection{Euclidean projection for system compatibility}

The final stage of our framework prepares the resized KB for use in original communication systems. Since most VQ-VAE-based models operate in Euclidean space, as noted in Section \ref{sec:system_model}, we must project the pruned set of hyperbolic vectors, $\mathcal{P}_K$, from the Poincaré ball back to a Euclidean space. This is achieved using the logarithmic map, which is the formal inverse of the exponential map at the origin (i.e., \eqref{eq:exp_map}). The transformation for each vector $\mathbf{p}_k \in \mathcal{P}_K$ is defined as:
\begin{equation}
    \label{eq:log_map}
    \mathbf{s}_{k} = \operatorname{arctanh}(\|\mathbf{p}_{k}\|) \frac{\mathbf{p}_{k}}{\|\mathbf{p}_{k}\|}, \text{for } k = 1, \dots, K.
\end{equation} This process yields the final, resized Euclidean KB, $\mathcal{S} = \{\mathbf{s}_k\}_{k=1}^{K}$, which is now fully compatible for integration with the digital communication system.

\begin{figure}
    \centering
    \includegraphics[width=0.85\linewidth]{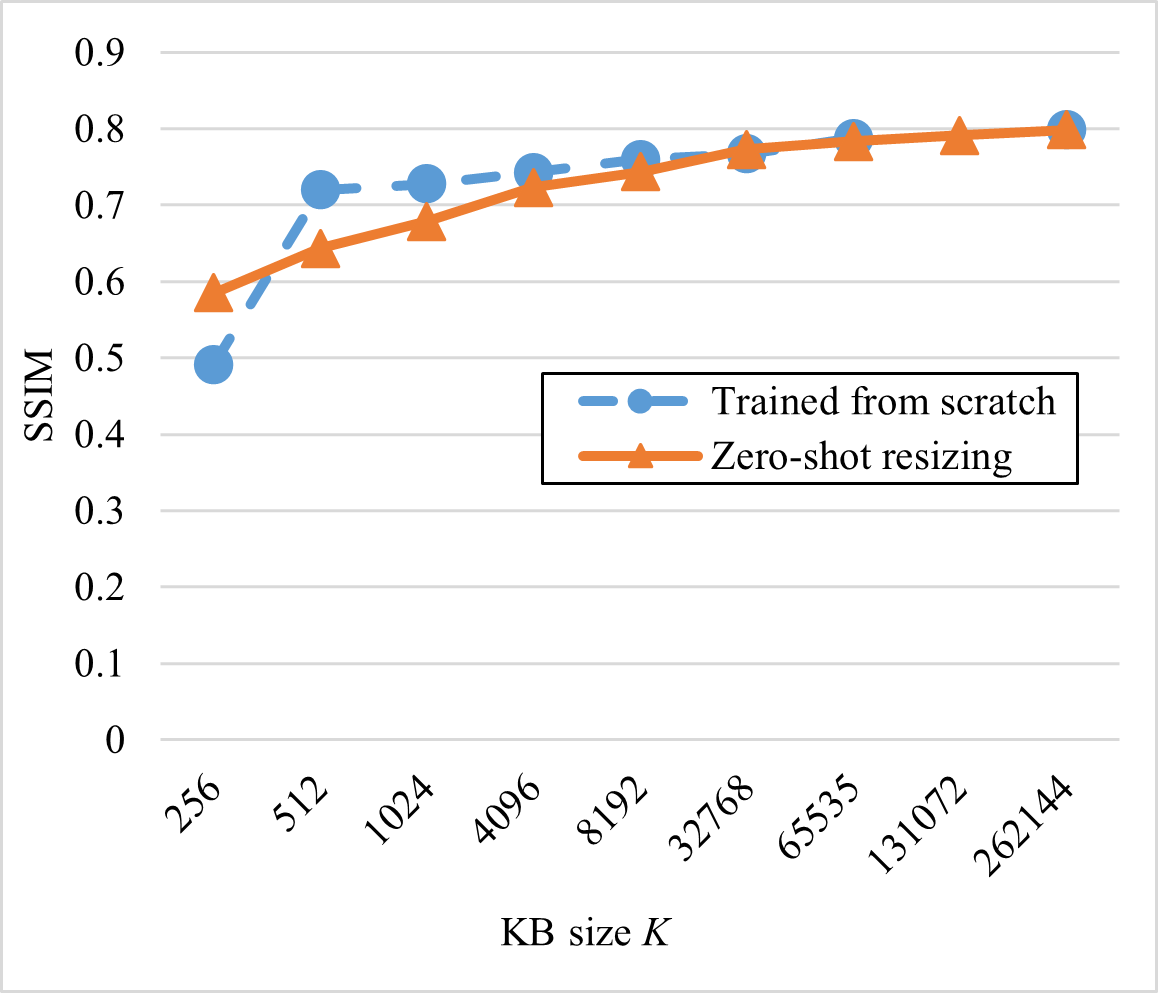}
    \caption{SSIM versus KB size $K$: Our zero-shot resizing method achieves nearly identical performance to the baseline of training dedicated KBs from scratch, without requiring retraining for each size.}
    \label{fig:SSIM}
\end{figure}

\begin{figure*}
    \centering
    \includegraphics[width=1\linewidth]{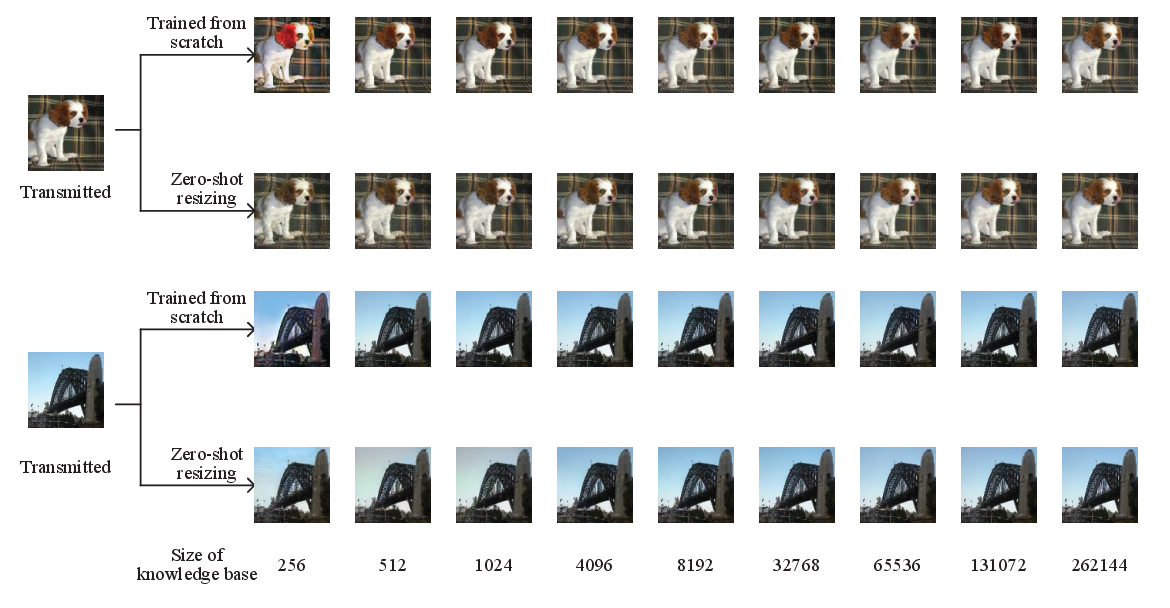}
    \caption{Qualitative comparison of reconstructed images under different KB sizes.}
    \label{fig:Visual}
\end{figure*}

\section{Evaluation} \label{sec:Exp}

In this section, we evaluate the performance of our proposed zero-shot KB resizing approach via extensive simulations. We conduct experiments on an end-to-end image semantic transmission task using the ImageNet dataset \cite{ImageNet}. Both the semantic encoder and decoder are built upon SimVQ \cite{SimVQ}, a state-of-the-art VQ-VAE framework that avoids representation collapse. We compare our approach against a baseline where a dedicated KB is individually trained from scratch for each size $K$. For clarity, these two methods are labeled ``Zero-shot resizing'' and ``Trained from scratch'' in the result figures, respectively. All simulations were conducted on a cluster of 8 NVIDIA V100 GPUs.

Fig.~\ref{fig:SSIM} plots the structural similarity index measure (SSIM) between source and reconstructed images under different KB size $K$. From this figure, we have the following two observations.
\begin{enumerate}
    \item SSIMs in both our design and the compared case increase with $K$, confirming that the expected behavior that a larger KB generally leads to higher reconstruction quality. 
    \item More importantly, SSIM of our design closely match with that of the individually trained baseline across all tested sizes. The average difference a negligible 0.8\%. This result is significant, as it demonstrates that our approach provides the flexibility of rate adaptation with almost no performance penalty compared to the computationally expensive method of training a separate, dedicated KB for each specific size.
\end{enumerate}

To provide a qualitative perspective, Fig.~\ref{fig:Visual} visualizes the reconstructed images of a sample from the test set. From this figure, we draw the following key observations:
\begin{enumerate}
    \item For any given KB size $K$, the image reconstructed by our zero-shot resizing method is visually almost identical to the one from the dedicated, trained-from-scratch model. This qualitative result strongly supports our quantitative observation from Fig.~\ref{fig:SSIM}, confirming that our resizing approach achieves comparable performance to the computationally intensive baseline without any retraining.
    \item A crucial advantage of our method is revealed at the lowest rate ($K=256$). The baseline model trained from scratch experiences a catastrophic failure, producing an image with severe artifacts and unnatural colors. This suggests that training a VQ-VAE-based end-to-end semantic communication model with a very small KB is inherently unstable. In stark contrast, our method maintains semantic coherence, producing a recognizable, albeit blurry, image. This demonstrates the robustness of our hierarchical pruning: by preserving the most foundational concepts from the large parent KB, our approach ensures a graceful degradation in quality rather than a complete collapse.
\end{enumerate}

Note that a major advantage of our approach is the dramatic reduction in training overhead. The ``Trained from scratch'' baseline requires training a new, dedicated model for each of the nine KB sizes evaluated in our experiments. In contrast, our ``Zero-shot resizing'' method only requires training a single parent model (with $K=262,144$). The subsequent resizing via hierarchical pruning is a computationally inexpensive post-processing step that does not require any further training or fine-tuning. By reducing the number of required training runs from nine to one, our approach achieves an approximate $9\times$ reduction in total training time and GPU resources. Therefore, our method not only provides superior flexibility and robustness but does so while demanding only a fraction of the computational budget required by the conventional approach.

\section{Conclusion} \label{sec:Conclusion}

In this paper, we addressed the critical challenge of rate adaptation in digital semantic communication by introducing a novel, zero-shot KB resizing method. The conventional approach of training dedicated KBs for different transmission rates is computationally prohibitive and lacks granularity. Our principled framework overcomes this by leveraging hyperbolic geometry to uncover the inherent semantic hierarchy within a single, large parent KB, enabling the instant creation of smaller, high-quality KBs of any arbitrary size without retraining. Our extensive evaluations confirmed that this approach yields reconstruction quality nearly identical to that of dedicated, from-scratch KBs while demanding only a fraction of the computational budget. Furthermore, our method demonstrates superior robustness with graceful performance degradation at very low rates, a scenario where conventional methods fail catastrophically. Ultimately, this work removes a significant architectural barrier, paving the way for truly flexible, efficient, and practical rate-adaptive semantic communication systems.

\section*{Acknowledgment}

This work is supported by the National Science and Technology Major Project - Mobile Information Networks under Grant No.2024ZD1300700, and by National Natural Science Foundation of China with grant 62301471.

\bibliographystyle{ieeetr}
\bibliography{ref}

\end{document}